\documentclass[preprintnumbers,amsmath,amssymb,nofootinbib,showpacs,onecolumn,a4paper]{revtex4}

\usepackage{graphicx}
\usepackage{bm}
\usepackage{amsmath}

\newcommand{\be}{\begin{equation}}
\newcommand{\ee}{\end{equation}}
\newcommand{\bea}{\begin{eqnarray}}
\newcommand{\eea}{\end{eqnarray}}
\newcommand{\bdm}{\begin{displaymath}}
\newcommand{\edm}{\end{displaymath}}
\newcommand{\beas}{\begin{eqnarray*}}
\newcommand{\eeas}{\end{eqnarray*}}

\newcommand{\mD}{\mathcal{D}}

\newcommand{\kv}{\mathbf{k}}

\newcommand{\xv}{\mathbf{x}}

\newcommand{\av}[1]{\left< #1\right>}
\newcommand{\dom}[1]{#1_{\mathcal{D}}}
\newcommand{\intdom}[1]{\frac{1}{\dom{V}}\int_{\mathcal{D}} #1\sqrt{h}d^3\mathbf{x}}
\newcommand{\mH}{\mathcal{H}}

\newcommand{\Vd}{V_{\mathcal{D}}}
\newcommand{\Vf}{V_{\mathcal{F}}}
\newcommand{\Rd}{\mathcal{R}_{\mathcal{D}}}
\newcommand{\Pd}{\mathcal{P}_{\mathcal{D}}}

\newcommand{\Qd}{\mathcal{Q}_{\mathcal{D}}}

\newcommand{\Fd}{\mathcal{F}_{\mathcal{D}}}
\newcommand{\Td}{\mathcal{T}_{\mathcal{D}}}

\newcommand{\Ld}{\mathcal{L}_{\mathcal{D}}}

\newcommand{\Rs}{R_\mathcal{S}}

\newcommand{\bkr}{\overline{\rho}}
\newcommand{\bkp}{\overline{p}}

\newcommand{\dtx}{\mathrm{d}^3\mathbf{x}}

\newcommand{\ktwoint}{\int k^2\mathcal{P}(k)\left|\phi_0(k)\right|^2\frac{{\rm d}k}{k}}
\newcommand{\kfourint}{\int k^4\mathcal{P}(k)\left|\phi_0(k)\right|^2\frac{{\rm d}k}{k}}
\newcommand{\kfourintw}{\int k^4\mathcal{P}(k)\left|\phi_0(k)\right|^2W^2(k\dom{R})\frac{{\rm d}k}{k}}

\begin{document}

\title{Backreaction: Gauge and Frame Dependences}

\author{Iain A. Brown}
\email{ibrown@astro.uio.no}
\affiliation{Institute of Theoretical Astrophysics, University of Oslo, P.O. Box 6094, N-0315 Blindern, Norway}
\author{Joey Latta}
\email{lattaj@mathstat.dal.ca}
\author{Alan Coley}
\email{aac@mathstat.dal.ca}
\affiliation{Department of Mathematics and Statistics, Dalhousie University, Halifax, Canada B3H 3J5}

\date{\today}

\begin{abstract}
The cosmological backreaction from perturbations is clearly gauge-dependent, and obviously depends on the choice of averaged Hubble rate. We consider two common choices of Hubble rate and advocate the use of comoving volume-preserving gauges. We highlight two examples valid to an appropriate order in perturbation theory, uniform curvature gauge, which is as close to volume-preserving as possible, and a spatially-traceless uniform cold dark matter gauge which preserves the volume to linear order. We demonstrate the strong gauge- and frame-dependences in averaging. In traceless uniform CDM gauge the backreaction exhibits a strong ultra-violet divergence and can be tuned to an arbitrary magnitude with an appropriate choice of smoothing scale. In uniform curvature gauge we find that for a choice of Hubble rate locked to the spatial surface the backreaction vanishes identically, while for a Hubble rate defined from a fluid's expansion scalar the effective energy density at the current epoch in an Einstein-de Sitter universe is $\Omega_{\rm eff}\approx 5\times 10^{-4}$, slightly bigger than but in broad agreement with previous results in conformal Newtonian gauge.
\end{abstract}

\maketitle

\section{Introduction}
\noindent The gravitational field equations on cosmological scales are obtained by averaging the Einstein field equations of general relativity. The effects of averaging (or backreaction) can have a significant dynamical effect on the evolution of the Universe and consequently on cosmological observations. The quantitative size of these effects, and their potential qualitative effect on cosmological observations, are currently of great interest (see, for example, \cite{BOOK}). However, in order for the results of cosmological averaging to make any physical sense, it is absolutely crucial to have a rigorous (fully covariant) definition of the spacetime average of a tensor on a differential manifold.

Cosmological perturbation theory provides a well-motivated paradigm in which to perform cosmological averaging. In perturbation theory we shall argue that a spacetime average is only well-defined when undertaken in so-called volume-preserving coordinates (VPC), or in the closely related comoving volume-preserving gauges (VPGs), which we define below. In fact, a VPG is ideally-suited to the study of spatial averaging in perturbation theory.

We discuss two such gauges which are valid to an appropriate order in perturbation theory; the uniform curvature gauge (which is uniquely well-defined in this respect and is as close to volume-preserving as possible) and the spatially-traceless uniform cold dark matter gauge (which is adequate for calculations to second order in perturbation theory). In fact, traceless gauges, including both of these, simplify the problem considerably since it reduces to an averaging of the product of linear perturbations and, at most, a second-order velocity.

An important measure of the effect of backreaction is the difference between the input Hubble rate and the averaged Hubble rate. We demonstrate that the cosmological backreaction from perturbations is strongly gauge-dependent, and clearly depends on the choice of frame for the averaged Hubble rate (e.g., the gravitational or the projected fluid frames). We shall show that in uniform curvature gauge the backreaction vanishes identically in the gravitational frame. We consequently argue that the definition of the Hubble rate should reference the fluid content of the universe and so it is necessary to use the projected fluid frame. We show that the backreaction in traceless uniform CDM gauge exhibits a strong dependence on choice of smoothing scale. In the projected fluid frame, the ultraviolet catastrophes in traceless uniform CDM gauge are exacerbated.

We then turn to some quantitative numerical results. We find that in uniform curvature gauge the effective energy density of backreaction at the current epoch in an Einstein-de Sitter universe is of order $10^{-4}-10^{-3}$, and is slightly larger (by a factor of 2-5) than, but in broad agreement with, previous results obtained in conformal Newtonian gauge. This slightly larger result perhaps suggests that backreaction within perturbation theory can be large enough to affect cosmological observations, but we believe the more important result is that this value is now on a significantly firmer basis than before. Of course, more significant quantitative results are possible in more realistic models in which the present-day universe is not well-described by perturbation theory.

\section{Formalism}
\noindent Three approaches to spatial averaging in perturbation theory and applicable in general coordinates have recently been presented \cite{Larena:2009,Brown:2009b,Gasperini:2009,Gasperini:2009mu,Umeh:2010pr}. The relationships between these have been explored in \cite{Larena:2009,Umeh:2010pr} and are shown to tend towards agreement for extremely large domains. The explicitly gauge-invariant approach of \cite{Gasperini:2009,Gasperini:2009mu} has been further developed and applied to averages across the past light-cone in \cite{Gasperini:2011us,BenDayan:2012pp}. However, our ultimate aim is to select a gauge in which the averaging procedure remains well-defined and so will focus on the formalisms of \cite{Larena:2009,Brown:2009b,Umeh:2010pr}. Likewise multiscale averaging models have been introduced (we see for instance \cite{Wiltshire07,Wiltshire:2009db,Wiegand:2010uh,Baumann:2010tm,Viaggiu:2012dg}) but for simplicity we consider only single-scale models.

A globally hyperbolic spacetime can be foliated with a family of 3-surfaces described by coordinates $x^\mu=(t,x^a)$. The 4-vector normal to the surfaces is $n^\mu=(1/\alpha)(1,-\beta^i)$ with normalisation $n^\mu n_\mu=-1$ with the lapse $\alpha$ and the shift $\beta^i$. The projection tensor onto the 3-surfaces, induced 3-metric and line element are then
\be
h_{\mu\nu}=g_{\mu\nu}+n_\mu n_\nu, \quad h_{ij}=g_{\mu\nu}h^\mu_ih^\nu_j,
\quad
\label{3+1LineElement}
ds^2=(-\alpha^2+\beta_i\beta^i)dt^2+2\beta_idtdx^i+h_{ij}dx^idx^j
\ee
where $\beta_i=h_{ij}\beta^j$. The embedding of the hypersurfaces is described by the extrinsic curvature $K_{ij}=-(1/2)\mathcal{L}_nh_{ij}=-(1/2)(\partial/\partial t)h_{ij}-2\mD_{(i}\beta_{j)}$. $\mD_i$ is the covariant derivative on the 3-surface and brackets denote symmetrisation on the enclosed indices.

Let $\mD$ be a finite domain lying on the inhomogeneous 3-surface, and let $h$ be the determinant of the 3-metric. The volume of this domain (see, e.g. \cite{Russ:1997,Buchert:2000,Buchert:2001,Wetterich:2003,Rasanen:2003}) and therefore the average over the volume of a scalar $A$ can be defined by
\be
\label{AverageDef}
\dom{V}=\int_{\mD}\sqrt{h}d^3\mathbf{x}, \quad \av{A}=\intdom{A}.
\ee

An average Hubble rate can be defined by
\be
\label{HubbleDef}
3\dom{\mH}=\frac{\dom{\dot{V}}}{\Vd}=\intdom{\left(\frac{1}{2}h^{ij}\dot{h}_{ij}\right)}
=\av{-\alpha K+\mD^i\beta_i}
\ee
with $\mD^i=h^{ij}\mD_j$. We dub this choice of Hubble rate the ``gravitational frame'' -- it describes the Hubble rate defined with respect to observers comoving with the coordinate grid.

It is possible instead to define an averaged Hubble rate from the expansion scalar of an observer, projected onto the spatial 3-surface, $\theta=h^{\mu\nu}u_{\mu;\nu}$. We term this the ``projected fluid frame'' and it was employed, for instance, in \cite{Larena:2009,Umeh:2010pr}. The Hubble rate is defined by
\be
3\dom{\mH}=\av{\alpha^2\theta} .
\ee
We should emphasise that this definition is not unique to this study and was introduced in \cite{Larena:2009}, and employed in \cite{Umeh:2010pr} to study the expansion scalar of a fluid tilted with respect to the averaging 3-surface -- that is, the expansion of the fluid as observed in the ``gravitational rest-frame''. We employ it in the same manner. Being defined from a physically-meaningful quantity this definition of the Hubble rate is perhaps to be preferred.

\section{Averaging in an Arbitrary Gauge}
\noindent The above equations are a rephrasing of the scalar 3+1 equations averaged across a 3D domain and as such provide no new information. To make progress either some behaviour (as in \cite{Larena:2008}), or some underlying model (as in for instance \cite{Rasanen:2004,Paranjape:2006,Rasanen:2008,Marra:2008sy,Behrend:2008,Brown:2009a,Clarkson:2009,Brown:2009cy,Umeh:2010pr}) must be assumed. Cosmological perturbation theory provides a well-motivated toy case. This approach introduces a local scale factor $a(t)$ and Hubble rate governing the unperturbed dynamics. One measure of the impact of ``backreaction'' is then to identify the backreaction as the difference between the input Hubble rate and the averaged Hubble rate. Of course, many other effects of backreaction will occur in cosmology \cite{BOOK}. In this paper, it is this Hubble-``backreaction'' that we are interested in.

The perturbed flat Robertson-Walker line element can be written
\be
\label{FLRWLineElement}
ds^2=a^2(\eta)\left(-(1+2\phi)d\eta^2
+2B_id\eta dx^i+\left(\delta_{ij}+2C_{ij}\right)dx^idx^j\right)
\ee
where $B_i=\partial_iB-S_i$ and $C_{ij}=-\psi\delta_{ij}+\partial_i\partial_jE+\partial_{(i}F_{j)}+\frac{1}{2}h_{ij}$ with $\partial^iS_i=\partial^iF_j=\partial^ih_{ij}=\delta^{ij}h_{ij}=0$. The fluid velocity is $v_i=\partial_iv+v_i^{(V)}$. Perturbations are expanded to second order with $\phi=\phi_{(1)}+(1/2)\phi_{(2)}$ and similar. We neglect everywhere products which are cubic or higher.\footnote{Note that this implies that if, for instance, the linear and second-order integrands are highly oscillatory, a dominant contribution could naturally arise at higher orders in perturbation theory. However, the study of such systems seems better suited to fully nonlinear relativistic models.} Indices are raised and lowered with the Kronecker delta. Derivatives with respect to conformal time will be denoted with an overdot and $\mH=\dot{a}/a$ is the conformal Hubble rate of the underlying model which obeys the Friedmann equations,
\be
\label{Background}
\mH^2=\frac{8\pi G}{3}a^2\sum_f\bkr_{(f)}+\frac{1}{3}a^2\Lambda,
\quad
\dot{\mH}=-\frac{4\pi G}{3}a^2\sum_f\left(\bkr_{(f)}+3\bkp_{(f)}\right)+\frac{1}{3}a^2\Lambda.
\ee
Linear tensor modes are extremely small and decay with the expansion of the universe while, unless supported by an active source, vector modes rapidly decay. To a first approximation, scalar modes therefore dominate the linear perturbations \cite{Durrer04}. At second-order all three types should in principle be considered, but it turns out that in the situations we study, the leading order contribution from vectors and tensors is quadratic, implying that to second-order in perturbations they can be neglected.

The metric determinant of the perturbed model in a general gauge is
\be
\sqrt{h}=1+C+\frac{1}{2}C^2-C^{ij}C_{ij} .
\ee
If $W(\mathbf{x})$ is a window function defining the domain, and
\be
V_\mathcal{F}=\int W(\mathbf{x})\sqrt{h_0(\mathbf{x})}\dtx=a^3\int W(\mathbf{x})\dtx
\ee
is the domain volume projected onto the FLRW background, the inhomogeneous domain volume is therefore
\be
\label{VolumeRelations}
\dom{V}=V_\mathcal{F}+a^3\int W(\mathbf{x})C(\mathbf{x})\dtx
+a^3\int W(\mathbf{x})\left(\frac{1}{2}C^2(\mathbf{x})-C^{ij}(\mathbf{x})C_{ij}(\mathbf{x})\right)\dtx
\nonumber
\ee
with the integration taken across all $\mathbf{x}$. When required to specify the window function we choose $W(y)=\exp(-y^2)$. Likewise, the average of a quantity $A(\mathbf{x})=A_1(\mathbf{x})+A_2(\mathbf{x})/2$ is 
\be
\av{A(\mathbf{x})}=\frac{a^3}{\Vd}\int W(\mathbf{x})A(\mathbf{x})\left(1+C(\mathbf{x})\right)\dtx
=\frac{a^3}{\Vd}\int W(\mathbf{x})\left(A_1(\mathbf{x})+\frac{1}{2}A_2(\mathbf{x})+A_1(\mathbf{x})C_1(\mathbf{x})\right)\dtx
\ee
while that of a product of linear perturbations $A_1(\mathbf{x})B_1(\mathbf{x})$ is
\be
\av{A_1(\mathbf{x})B_1(\mathbf{x})}=\frac{a^3}{\Vd}\int W(\mathbf{x})A_1(\mathbf{x})B_1(\mathbf{x})\dtx .
\ee
It is usual to then Taylor-expand the domain volume
\be
\Vd^{-1}=\Vf^{-1}\left(1-\frac{a^3}{\Vf}\int W(\xv)C(\xv)\dtx+\ldots\right)
\ee
where we truncate the expansion at linear order since corrections to averages of the type $\av{A(\xv)}$ would enter at cubic order or above.

Cosmological perturbations can be represented in Fourier space. Denoting an ensemble average with an overbar, the power spectrum of two linear perturbations is
\be
\overline{A_1(\kv)B_1^*(\mathbf{k'})}
=\frac{2\pi^2}{k^3}\mathcal{P}(k)A_1(k)B_1^*(k)(2\pi)^3\delta(\mathbf{k-k}')
\ee
with primordial power spectrum
\be
\mathcal{P}(k)=A_\star\left(\frac{k}{k_\star}\right)^{n_s-1} .
\ee
The ensemble average of the domain volume is then
\be
\overline{\dom{V}}=V_\mathcal{F}+a^3\overline{\int W(\mathbf{x})C(\mathbf{x})\dtx}
+\Vf\int\mathcal{P}(k)\left(\left|C(k)\right|^2-\frac{1}{2}C^{ij}(k)C^*_{ij}(k)\right)\frac{dk}{k} .
\ee
Likewise the ensemble average of a spatially-averaged product of linear perturbations becomes
\be
\label{EnsembleAverageLinearProduct}
\overline{\av{A_1(\mathbf{x})B_1(\mathbf{x})}}=\frac{1}{2}\int\mathcal{P}(k)\left(A_1(k)B^*_1(k)+\mathrm{c.c.}\right)\frac{dk}{k} .
\ee
The ensemble average of a linear perturbation is vanishing by definition.

In the gravitational frame, the averaged Hubble rate (\ref{HubbleDef}) then becomes
\be
\label{PerturbedHubble}
\dom{\mH}=\mH+\frac{1}{3}\av{\dot{C}-2C^{ij}\dot{C}_{ij}} .
\ee
This corresponds with the large-scale limit of the the expansion scalar of the coordinate grid defined with respect to conformal time \cite{Malik:2008}, $\theta_{\rm conf}\approx 3\left(\mH-\dot{\psi}-2\psi\dot{\psi}\right)$. The perturbed Hubble rate in the backreaction is a simple average of this quantity (denoted $\xi$ in \cite{Umeh:2010pr}).

Squaring this Hubble rate gives
\be
\dom{\mH}^2=\mH^2+\frac{2}{3}\mH\av{\dot{C}-2C^{ij}\dot{C}_{ij}}+\frac{1}{9}\av{\dot{C}}^2
\ee
and so the effective energy density
\be
\mH^2\Omega_{\rm eff}=\frac{2}{3}\mH\av{\dot{C}-2C^{ij}\dot{C}_{ij}}+\frac{1}{9}\av{\dot{C}}^2 .
\ee
It is important to note that in the gravitational frame the \emph{entire} backreaction then depends solely on the choice of threading of the 3-surface, defined by the choice of $C_{ij}$. The choice of slicing (i.e. of $\phi$ and $B_i$) and the behaviour of any fluid content influences $\dom{\mH}$ only indirectly through dynamics.

In the projected fluid frame the averaged Hubble rate is instead
\be
\label{FluidFrameHubbleRate}
\dom{\mH}=\mH+\frac{1}{3}\av{\partial^i\partial_iv+\dot{C}+2\mH C}
+\frac{1}{3}\Big<\phi\partial^i\partial_iv+\partial_iv\partial^iC-\partial_iB\partial^i\phi
+\frac{3}{2}\mH\partial_aV\partial^aV+2C^{ij}\partial_i\partial_jv -2\dot{C}^{ij}C_{ij}\Big>
\ee
and so depends both on the coordinates on the 3-surface, on the choice of slicing, and on the velocity of the fluid with respect to the background, $v$. The covariant velocity is $V=v+B$.

Both these forms of the Hubble rate are trivially gauge-dependent. This is not surprising. The choice of gauge has governed our choice of 3-surface upon which to average and we should not expect gauge-invariance. Rather we should attempt to find in which gauges an average can be properly defined.

\section{Gauge Choices}
\noindent Coordinate freedoms allow us to eliminate two scalar degrees of freedom. Studies of backreaction have typically been in synchronous gauge with $\phi=B=0$ (e.g.  \cite{Russ:1997,Buchert:2001,Li:2007}) or conformal Newtonian gauge with $B=E=0$ (e.g. \cite{Kasai:2006,Tanaka:2007,Behrend:2008,Brown:2009a,Clarkson:2009,Brown:2009cy,Umeh:2010pr}). Uniform curvature gauge was proposed in \cite{Brown:2009cy} but not examined in detail, although in a different formalism \cite{Paranjape:2009} a volume-preserving system with similarities to uniform curvature gauge was considered.

In order to carry out any cosmological averaging, it is absolutely crucial to have a rigorous (fully covariant) definition of the spacetime average of a tensor on a differential manifold in order for the results to make any physical sense. There have been a number of recent approaches to this, including the exact macroscopic gravity (MG) approach which gives a prescription for the correlation functions that emerge in an averaging of the Einstein field equations \cite{Zalaletdinov:1997,Coley:2005,Coley:2006,Coley:2007,VanDenHoogen:2007en,VanDenHoogen:2009,Clifton:2012fs,BHC}) (see also \cite{BOOK}).

This is an absolute necessity for any results to be interpreted physically. For an explicit example consider the situation in MG. This approach rests on the definition of a bivector with specific properties; when one examines these, the only allowed coordinate systems are those which are volume-preserving \cite{Mars:1997jy,Paranjape:2009}. This is a generic feature of covariant averaging schemes. As a consequence, in the context of this paper -- averaging in perturbation theory -- in general a spacetime average is only meaningful when undertaken in a volume-preserving coordinate system (VPC). 

In a VPC the volume of a domain is preserved as the system evolves in time. Closely related are what we term comoving volume-preserving gauges (VPGs) in which the volume of a 3-domain on an inhomogeneous surface evolves purely as $a^3(\eta)$. In a VPG the time dependences cancel out when one takes an average -- for the purposes of spatial averaging, a VPG is then effectively a VPC.\footnote{A similar argument was employed in \cite{Paranjape:2008}} VPCs are employed in, for instance, Macroscopic Gravity \cite{Zalaletdinov:1997,Coley:2005,Coley:2006,Coley:2007,VanDenHoogen:2007en,VanDenHoogen:2009,Clifton:2012fs} and a VPC at linear order was applied to cosmological averaging in \cite{Paranjape:2009}. VPCs are also explicitly utilized in an approach to averaging within unimodular gravity \cite{Unimod}. It is a central aim of this paper to study the significance of the VPGs, or whether it is adequate to average in another, practically convenient gauge.

Written explicitly, the spatial average of a perturbation $A(\xv)$ in arbitrary coordinates is given by
\be
\av{A(\xv)}=\frac{\int W(\xv)A(\xv)\dtx+\int W(\xv)A(\xv)C(\xv)\dtx}{\int W(\xv)\dtx+\int W(\xv)C(\xv)\dtx+\int W(\xv)\left(\frac{1}{2}C^2(\xv)-C^{ij}(\xv)C_{ij}(\xv)\right)\dtx} .
\ee
This takes on the simplest form when $C_{ij}=0$. Tensor modes at linear-order are gauge-invariant and cannot be removed, but the gauge that imposes $C_{ij}=(1/2)h_{ij}$ at an arbitrary order in perturbation theory is uniform curvature gauge. Neglecting tensor perturbations,
\be
\Vd=a^3\int W(\xv)\dtx, \quad \av{A(\xv)}=\frac{\int W(\xv)A(\xv)\dtx}{\int W(\xv)\dtx} .
\ee
This is then a VPG; the spatial surfaces align with the FLRW background, and the 3-volume expands only with the background and the only time-dependence in the average is that of the perturbation itself. This gauge is ideally-suited to the study of spatial averaging in perturbation theory. The tensor modes can be included. To second order, tensor and scalar contributions do not couple together, and the results of \cite{Brown:2009cy} can be directly employed: tensor modes from inflation will produce a baseline backreaction of order $\Omega_{\rm eff}\approx 10^{-14}$.

Uniform curvature gauge is the unique choice for a comoving VPG, but a less stringent alternative valid to second-order when averaging perturbations can also be found. If we assume that we can Taylor-expand $\Vd^{-1}$, we can reduce the spatial average to
\be
\label{TracelessAverage}
\av{A(\xv)}\frac{\left(a^3\int W(\xv)A(\xv)\dtx+a^3\int W(\xv)A(\xv)C(\xv)\dtx\right)\left(1-\dfrac{a^3}{\Vf}\int C(\xv)\dtx\right)}{\Vf}
\ee
where we have truncated to second-order in perturbations. From this it is clear that if we choose a (spatially) traceless gauge, with $C(\xv)=0$, the corrections to the volume are pushed to higher-orders in perturbation theory. Uniform curvature gauge is a trivial example of a traceless gauge, but a convenient, less stringent gauge can be found which is sufficiently close to a VPG and may provide a simpler basis for calculations than uniform curvature gauge.

A traceless gauge is defined by
\be
C_T=0 \Rightarrow 3\psi_T=\partial^a\partial_aE_T .
\ee
The scalar transformation from an arbitrary gauge (written with a tilde) into this gauge is given by a generating vector $\xi^\mu=(\alpha,\partial^i\beta)$ with
\be
\begin{array}{c}
\phi_T=\tilde{\phi}+\dot{\alpha}+\mH\alpha,
\quad
C_T=\tilde{C}+3\mH\alpha+\partial^a\partial_a\beta=0
\\
B_T=\tilde{B}-\alpha+\dot{\beta},
\delta_T=\tilde{\delta}-3\mH(1+w)\alpha,
\quad
v_T=\tilde{v}-\dot{\beta} .
\end{array}
\ee
A choice containing no arbitrary constant is a uniform density gauge, where the spatial sections follow contours of constant density of some fluid. Aligning to CDM gives the generating vector
\be
\alpha=\frac{\tilde{\delta}_c}{3\mH}, \quad \partial^a\partial_a\beta=3\tilde{\psi}-\tilde{\delta}_c+\partial^a\partial_a\tilde{E} .
\ee
In terms of Newtonian gauge quantities the traceless gauge perturbations are therefore
\be
\label{TracelessTransformations}
\begin{array}{c}
\psi_T=\psi_N-\dfrac{\delta_{cN}}{3}, \quad
E_T=-\dfrac{3}{k^2}\psi_T, \quad
\phi_T=\phi_N+\dfrac{1}{\mH}\left(\dot{\psi}_N+\dfrac{1}{3}k^2v_{cN}-\dfrac{1}{3}\dfrac{\dot{\mH}}{\mH}\delta_{cN}+\delta_{cN}\right), \\
B_T=v_{cN}-\dfrac{1}{3\mH}\delta_{cN}, \quad
\delta_{aT}=\delta_{aN}-(1+w_A)\delta_{cN}, \quad
v_{aT}=v_{aN}-v_{cN}
\end{array}
\ee
where we have used that in Newtonian gauge $\dot{\delta}_{cN}=3\dot{\psi}_N+k^2v_{cN}$. Note importantly that this gauge is therefore a uniform CDM density gauge comoving with CDM.\footnote{It can be motivated as such; setting $v_{cT}=\delta_{cT}=0$ one recovers the above gauge with an arbitrary function of $\xv$ appearing in $E_T$. Since this constant only influences the coordinates on the 3-surface and not the choice of slicing it can be fixed to ensure $C(\xv)=0$.}

It is interesting to note that the use of these gauges simplifies the evaluation of backreaction in an another important, practical manner: it removes inconvenient terms. Consider the averaged Hubble rate in the gravitational frame. This then contains the average of $\dot{C}$. In an arbitrary gauge this contains spatial averages of both linear perturbations and second-order perturbations. While the ensemble averages of linear perturbations vanish, this still leaves an average across a second-order perturbation and products of spatial averages $\av{\dot{C}_1}\av{C_1}$. However, we only have a firm method for evaluating the averages of products of linear perturbations. The second-order term is particularly problematic. Approximations for the nonlinear Bardeen potential in $\Lambda$CDM exist \cite{Bartolo:2005kv,Clarkson:2009,Umeh:2010pr} and so, in principle, this term can be calculated in all gauges. These solutions are, however, valid only in matter-dominated universes (with or without a cosmological constant), and it would obviously be convenient to have a formalism that can be readily applied to universes with arbitrary fluid content. Choosing a traceless gauge immediately removes this term identically, and the form of the Hubble rate becomes significantly simpler.

Similarly, in the projected fluid frame we must average across $C(\xv)$ and $\partial^a\partial_a v$, and a traceless comoving gauge would appear to render the problem sufficiently straightforward. In uniform curvature gauge we still require knowledge of $v_{2F}$. The desire, therefore, is that the almost-volume preserving traceless uniform CDM gauge can provide an ``accurate-enough'' approximation to the result in the genuinely volume-preserving uniform curvature gauge.

In this manner one can motivate the choices of gauge we wish to study. Averaging in uniform curvature gauge is uniquely well-defined, and we consider it in both the gravitational and the projected fluid frames. However, the traceless uniform CDM gauge is also well-motivated and preserves the comoving volume up to linear order, and averages up to second order. In the gravitational frame, any traceless gauge -- including both uniform curvature and traceless uniform CDM -- reduces the problem to averaging the product of linear perturbations. In contrast, in the projected fluid frame a traceless gauge comoving with the fluid provides the most convenient choice. We will also consider Newtonian gauge in the gravitational frame to provide a direct comparison with previous results \cite{Brown:2009a,Clarkson:2009,Umeh:2010pr,Kolb:2011zz}.

As a final comment, we have been assuming that the corrections to $\Vd$ are small, and that we can take a Taylor expansion of the inverse volume. Physically one would not expect an integral across a gravitational potential to give a large result, but the correction integrals in the volume are frequently divergent in the infra-red, and potentially also in the ultra-violet (see for instance the power spectra in \cite{Wands:2009ex,Yoo:2010ni,Bruni:2011ta}; calculations of the volume reduce to integrals across such spectra). These divergences contribute to the spatial averages, and could render the Taylor series physically valid, but mathematically suspicious. Since it is physically implausible that metric perturbations cause the volume of a domain to become arbitrarily large we follow standard procedure and neglect this issue. Nevertheless, it is worth noting that the only gauge in which this does not arise is the uniform curvature gauge in which $\Vd=\Vf$, neglecting tensor modes. Preserving the consistency of the formalism may then demand that we work in this gauge.

\section{Dynamics}
\noindent Up to this point our treatment has been applicable to any perturbed flat FLRW universe. For simplicity\footnote{And to ease direct comparison with previous work.} we choose henceforward to work in a pure dust Einstein-de Sitter universe with $\Omega_m=1$, $\Omega_\Lambda=0$ and $\mH=2/\eta$, unless stated otherwise. When necessary we choose $\Omega_b=0.05$ and $h=0.704$. The amplitude of primordial perturbations will be $\mathcal{A}_\star^2=2.42\times 10^{-9}$, and we employ a Harrison-Zel'dovich spectrum with $n_s=1$ for simplicity. A small spectral tilt will not significantly change our answers.

This model serves as a reasonable approximation up until recent redshifts. Analytic solutions for the Newtonian potentials in a dust-dominated universe (with or without a cosmological constant) at both linear and second-order were derived in \cite{Bartolo:2005kv} and applied to the averaging problem in \cite{Clarkson:2009,Umeh:2010pr}, and we quote the results directly here.

In Newtonian gauge the linear potentials, fluid density contrast and velocity are given by
\be
\phi_{1N}=\psi_{1N}=g(\eta)\phi_0(\xv), \quad \delta_{1N}=-2\phi_{1N}+\frac{1}{6}\eta^2\partial^a\partial_a\phi_{1N}-\eta\dot{\phi}_{1N}, \quad v_{1N}=-\frac{1}{3}\eta\phi_{1N}-\frac{1}{6}\eta^2\dot{\phi}_{1N}
\ee
where $g(\eta)$ is the growth function and $\phi_0$ the value of the Newtonian potential at the present epoch. At second-order, the potentials are given by
\bea
\psi_{2N}&=&A_1(\eta)\phi_0^2+A_2(\eta)\chi_2(\phi_0)+A_3(\eta)\chi_3(\phi_0)+A_4(\eta)\partial^i\phi_0\partial_i\phi_0, \\
\phi_{2N}&=&\tilde{A}_1(\eta)\phi_0^2+\tilde{A}_2(\eta)\chi_2(\phi_0)+\tilde{A}_3(\eta)\chi_3(\phi_0)+\tilde{A}_4(\eta)\partial^i\phi_0\partial_i\phi_0
\eea
with $\phi_{2N}\neq\psi_{2N}$ due to an effective anisotropic stress arising from products of linear perturbations. Here $A_n(\eta)$ and $\tilde{A}_n(\eta)$ are functions of time related to the expansion of the background which can be found in \cite{Bartolo:2005kv}, while $\chi_n(\eta)$ are second-order products of gradients and inverse Laplacians of $\phi_0$,
\be
\chi_2(\phi_0)=\partial^{-2}\left(\partial^i\phi_0\partial_i\phi_0\right)-3\partial^{-4}\partial_i\partial^j\left(\partial^i\phi_0\partial_j\phi_0\right), \quad
\chi_3(\phi_0)=\partial^{-2}\partial_i\partial^j\left(\partial^i\phi_0\partial_j\phi_0\right) .
\ee
To find the velocity in uniform curvature gauge we also require the Laplacian of the scalar velocity, which can be found from the momentum constraint \cite{Christopherson:2011ra},
\bea
\lefteqn{
4\pi Ga^2(\rho+p)\partial^i\partial_iv_{2N}
=-\partial^i\partial_i\dot{\psi}_{2N}-\mH\partial^i\partial_i\phi_{2N}+8\pi Ga^2(\rho+p)\partial^i\left(\phi_{1N}\partial_iv_{1N}\right)+16\pi Ga^2\partial^i\left(\psi_{1N}\partial_iv_{1N}\right)
}
\\ && \quad
-8\pi Ga^2\partial^i\left[(\delta\rho_{1N}+\delta p_{1N})\partial_iv_{1N}\right]
-4\partial^i\left(\psi_{1N}\psi_{1N}\right)^\cdot+2\partial^i\left[\partial_i\phi_{1N}\left(\dot{\psi}_{1N}+4\mH\phi_{1N}\right)\right]+4\partial^i\left(\phi_{1N}\partial_i\dot{\psi}_{1N}\right) .
\nonumber
\eea
In EdS $g(\eta)=1$ and the solutions reduce to
\bea
\label{AnalyticPsi1}
\psi_{1N}=\phi_{1N}=\phi_0, \quad \delta_{1N}=-2\phi_0^2+\frac{1}{6}\eta^2\partial^a\partial_a\phi_0, \quad v_{1N}=-\frac{1}{3}\eta\phi_0, \\
\label{AnalyticPsi2}
\left.\begin{array}{rl}
\label{AnalyticPhi2}
\psi_{2N}=-2\phi_0^2-\dfrac{4}{3}\chi_2(\phi_0)+B_3(\eta)\left(\chi_3(\phi_0)-\dfrac{3}{10}\partial^i\phi_0\partial_i\phi_0\right), \\
\phi_{2N}=2\phi_0^2+\dfrac{3}{2}\chi_2(\phi_0)+B_3(\eta)\left(\chi_3(\phi_0)-\dfrac{3}{10}\partial^i\phi_0\partial_i\phi_0\right)
\end{array}\right\}\Rightarrow \dot{\psi}_{2N}=\dot{\phi}_{2N}
\eea
with
\be
B_3(\eta)=\frac{2}{3}\eta^2\left(\frac{5}{14}-\frac{1}{2}\left(\frac{\eta_m}{\eta}\right)^2+\frac{1}{7}\left(\frac{\eta_m}{\eta}\right)^7\right)\approx \frac{5}{21}\eta^2 .
\ee
Here $\eta_m$ is an early time deep in matter domination at which the Newtonian potentials are initialised and we focus on the regime $\eta\gg\eta_m$. $\phi_0$ can be readily recovered from a Boltzmann code. The divergence of the velocity potential is then
\be
\frac{1}{\eta}\partial^a\partial_av_{2N}=\frac{1}{21}\eta^2\partial^i\partial_i\left(\partial^a\phi_0\partial_a\phi_0\right)-\frac{10}{63}\eta^2\partial^i\partial_j\left(\partial_i\phi_0\partial^j\phi_0\right)-\frac{8}{3}\partial^i\phi_0\partial_i\phi_0+\frac{1}{9}\eta^2\partial^i\left(\partial_i\phi_0\partial^a\partial_a\phi_0\right)+2\chi_3(\phi_0) .
\ee

The linear gauge transformation from Newtonian to uniform curvature gauge is generated by the 4-vector
\be
\xi_1^\mu=(\alpha,\partial^i\beta)=\left(\frac{\psi_{1N}}{\mH},\mathbf{0}\right)=\left(\frac{1}{2}\eta\phi_0,\mathbf{0}\right) .
\ee
With this transformation vector the linear uniform curvature quantities are readily found to be
\be
\phi_{1F}=\frac{5}{2}\phi_0, \quad
B_{1F}=-\frac{1}{2}\eta\phi_0, \quad
\delta_{1F}=-\left(5+\frac{1}{6}k^2\eta^2\right)\phi_0, \quad
v_{1F}=-\frac{1}{3}\eta\phi_0, \quad
V_{1F}=-\frac{5}{6}\eta\phi_0 .
\ee
The gauge transformation for the second-order velocity potential can be written \cite{Malik:2008} as
\be
\partial^a\partial_av_{2F}=\partial^a\partial_av_{2N}-\partial^a\partial_a\dot{\beta}_2+\partial^k\chi^v_k
\ee
with
\be
\chi^v_i=-2\alpha_1\partial_i\left(\dot{v}_{1N}+\mH v_{1N}\right)=\eta\phi_0\partial_i\phi_0, \quad
\beta_2=-\frac{3}{4}\partial^{-4}\partial_i\partial_j\chi^{ij}+\frac{1}{4}\partial^{-2}\chi^k_k ,
\ee
and the gauge function $\chi_{ij}$ is
\be
\chi_{ij}=-\frac{2}{\mH}\psi_{1N}\left(\dot{\psi}_{1N}+2\mH\psi_{1N}\right)\delta_{ij}-\frac{2}{\mH^2}\partial_i\psi_{1N}\partial_j\psi_{1N}
=-4\psi_{1N}^2\delta_{ij}-\frac{1}{2}\eta^2\partial_i\psi_{1N}\partial_j\psi_{1N} .
\ee
After some manipulation this gives the uniform curvature gauge velocity in a pure dust universe as
\begin{align}
\label{vf2}
\partial^a\partial_av_{2F}=
\frac{1}{21}\eta^3\partial^i\partial_i\left(\partial^a\phi_0\partial_a\phi_0\right)
-\frac{10}{63}\eta^3&\partial^i\partial_j\left(\partial_i\phi_0\partial^j\phi_0\right)
+\frac{1}{9}\eta^3\partial^i\left(\partial_i\phi_0\partial^a\partial_a\phi_0\right)
\nonumber \\
&+\frac{5}{4}\eta\chi_3(\phi_0)
+\eta\phi_0\partial^a\partial_a\phi_0
-\frac{17}{12}\eta\partial^i\phi_0\partial_i\phi_0
.
\end{align}

Finally, we need the curvature perturbation, lapse and shift in traceless uniform CDM gauge. From equations (\ref{TracelessTransformations}) these are
\be
\label{TracelessCDM}
\psi_{1T}=\frac{5}{3}\phi_0-\frac{1}{18}\eta^2\partial^a\partial_a\phi_0,
\quad
\phi_{1T}=\frac{5}{36}\eta^2\partial^a\partial_a\phi_0, \quad B_{1T}=-\frac{1}{36}\eta^3\partial^a\partial_a\phi_0 .
\ee

\section{The Gravitational Frame}
\subsection{Uniform Curvature Gauge}
\noindent A first step towards the use of uniform curvature gauge in this frame was presented in \cite{Brown:2009b}. However, in that study the authors only presented general forms and did not interpret the results. From equation (\ref{PerturbedHubble}), the averaged Hubble rate and effective energy density in uniform curvature gauge are then
\be
\dom{\mH}=\mH, \quad \mH^2\Omega_{\rm eff}=0.
\ee
The backreaction in uniform curvature gauge is identically zero, to an arbitrary order in perturbation theory! This result is contrary to the claim in \cite{Brown:2009b} that a gauge cannot be found that removes the backreaction.\footnote{In that study the authors did not address the effective energy density and instead demanded that each individual backreaction term in the ``Buchert'' approach vanish, and a gauge cannot be found in which that is the case. A gauge can, however, be found in which the combination of the backreaction terms vanishes, which we explicitly demonstrate in Appendix \ref{Buchert}.} This result was previously shown in \cite{Finelli:2011cw} in the context of cosmological inflation and assuming a long-wavelength limit;\footnote{We are grateful to an anonymous referee for drawing this result to our attention.} our treatment here is valid on all scales addressable with perturbation theory.

Presented in this manner, this result is trivial: the impact of scalar and vector perturbations on a Hubble rate in the gravitational frame vanishes identically, because we are working in a gauge with vanishing spatial scalar and vector perturbations. It is straightforward to interpret this result: the averaged Hubble rate in the gravitational frame is defined by the change in the volume of the domain. Since we are in a comoving volume-preserving gauge, the volume expands only with $a^3$, and so the averaged Hubble rate is given purely by the input Hubble rate.

This vanishing answer contradicts expectation -- while certainly one might argue the backreaction (that is, the ``Hubble-backreaction'') from perturbations should be small it cannot be expected to be identically zero at second- or higher-orders in perturbation theory. Given that uniform curvature gauge provides the best-motivated system in which to average, this suggests that the gravitational frame is ill-suited to studies of backreaction. Certainly it is difficult to connect the Hubble rate averaged in this frame with any physical quantity.

\begin{figure}
\includegraphics[width=0.9\columnwidth]{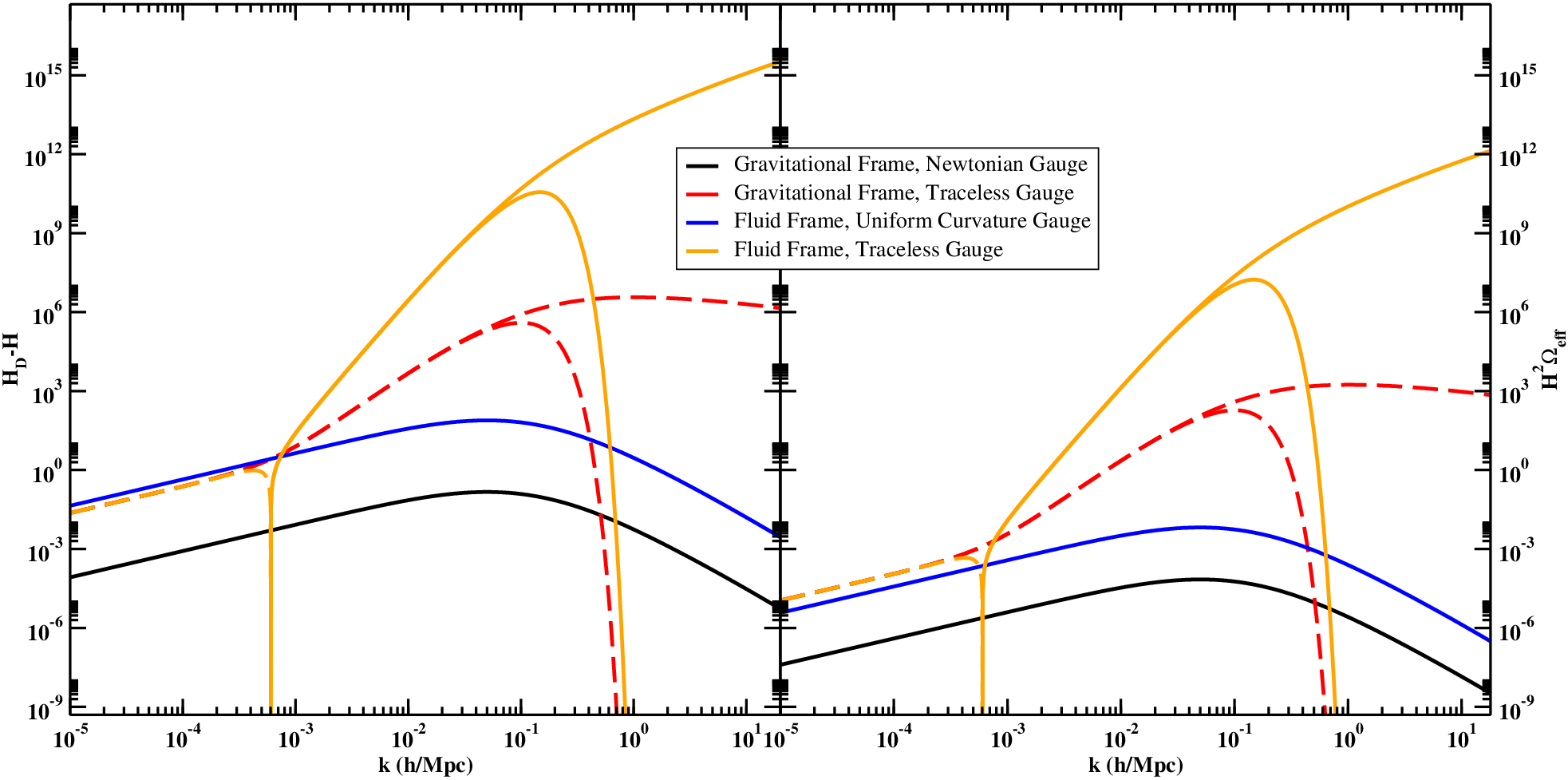}
\caption{Integrands for $\dom{\mH}-\mH$ (left) and $\mH^2\Omega_{\rm eff}$ (right). Dashed lines are negative.}
\label{Figure:Integrands}
\end{figure}

\subsection{Traceless Uniform CDM Gauge}
\noindent The above conclusions follow trivially from the definition of the Hubble rate in uniform curvature gauge, but as argued earlier we can relax the gauge constraint slightly and employ a traceless gauge. In the traceless uniform CDM gauge, the averaged Hubble rate and effective energy density become
\be
\overline{\dom{\mH}}=\mH-\frac{2}{3}\overline{\av{C^{ij}\dot{C}_{ij}}}, \quad
\mH^2\Omega_{\rm eff}=-\frac{4}{3}\mH\overline{\av{C^{ij}\dot{C}_{ij}}} .
\ee
Since $C_{ij}=-\psi_T\delta_{ij}+\partial_i\partial_jE_T$ and $3\psi_T=\partial^a\partial_aE_T$, the ensemble averages can be rewritten using equations (\ref{EnsembleAverageLinearProduct}) to become
\bea
\overline{\dom{\mH}}&=&\mH-\frac{20}{27}\eta\ktwoint-\frac{2}{81}\eta^3\kfourint, \\
\mH^2\Omega_{\rm eff}&=&-\frac{80}{27}\ktwoint-\frac{8}{81}\eta^2\kfourint .
\eea
We have written the curvature perturbation in this gauge in terms of the Newtonian gauge quantities using the gauge transform (\ref{TracelessCDM}).

We recover $\phi_0$ from a modified version of the \textsc{CMBFast} code \cite{Seljak:1996} itself based on \textsc{Cosmics} \cite{Bertschinger:1995er}, but we can gain insight examining the zero baryon transfer function found in \cite{Eisenstein:1997ik},
\be
\label{EisensteinHu}
T_0=\frac{L_0}{L_0+C_0q^2}, \quad q=k\;{\rm Mpc}\;h^{-1}\Theta_{\rm 2.7}^2/\Gamma, \quad \Gamma=\Omega_0h=h, \quad L_0=\ln(2e+1.8q), \quad C_0=14.2+\frac{731}{1+62.5q} .
\ee
This provides a good approximation to the numeric $\phi_0$; with $\Omega_b=0.05$ the baryon oscillations and small-scale damping from the baryons is relatively minor. The analytic form shows that there is an ultra-violet divergence in the term proportional to $k^4$; on large scales the integrand scales as $\sim(\ln k)^2/k$ which produces a logarithmic divergence $\sim(\ln k)^3/3$. We control this as in \cite{Clarkson:2009}, smoothing the gravitational potential in real space with the window function $W(x/\Rs)$. The smoothing scale $\Rs$ is arbitrary. A well-motivated choice would be the Silk scale, $\Rs=(k_{\rm Silk})^{-1}\approx 6{\rm Mpc}$ where the numerical value is for an Einstein-de Sitter universe.

Figure \ref{Figure:Integrands} shows the integrands, generated by combinations of the Newtonian gauge gravitational potential, which dominates on large scales, and the Newtonian gauge density contrast, which dominates on smaller scales. It is clear that the integral does not contain an infra-red divergence, and the integrand with and without the small-scale smoothing is plotted, taking $\Rs=6{\rm Mpc}$.

We solve the integrals numerically. For $\Rs=R_{\rm Silk}$ we have
\be
\frac{\overline{\dom{\mH}}-\mH}{\mH}=-0.409, \quad \Omega_{\rm eff}=-0.818
\ee
(Due to the additional small-scale damping in the baryon case the results found from the Eisenstein and Hu zero-baryon transfer function (\ref{EisensteinHu}) are slightly larger: $(\overline{\dom{\mH}}-\mH)/\mH=-0.511$, $\Omega_{\rm eff}=-1.02$.) The unphysically large size of these results strongly suggests that in this gauge the perturbations must be controlled on a larger scale.

\begin{figure}
\includegraphics[width=0.9\columnwidth]{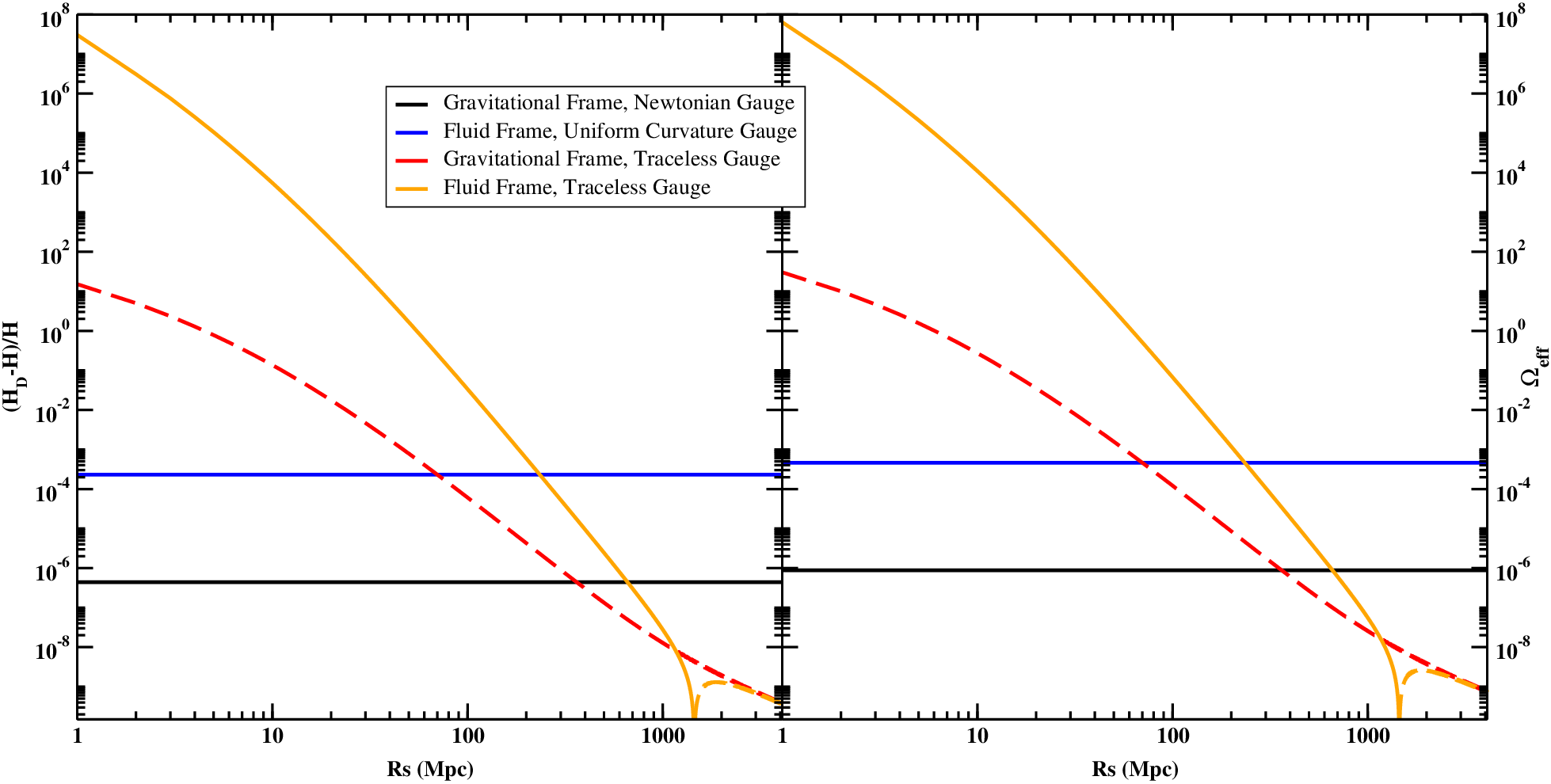}
\caption{Modification to Hubble rate (left) and effective energy density (right) at the current epoch as a function of smoothing scale $\Rs$. The domain radius is $\dom{R}=\eta_0/2$.}
\label{Figure:Solutions}
\end{figure}

The left panel of Figure \ref{Figure:Solutions} shows the fractional
shift $(\overline{\dom{\mH}}-\mH)/\mH$ as a function of smoothing scale
$\Rs$ for the test Einstein-de Sitter universe. The dependence is
extremely strong. For $\Rs\lesssim 4{\rm Mpc}$ the fractional change to the Hubble rate is significantly larger than unity. The magnitude of the correction decays monotonically as $\Rs\rightarrow\infty$; since the entire integral is smoothed there is no asymptote. It is important to note that while there is a dependence on the smoothing scale $\Rs$, there is no dependence on the averaging scale $\Rd$.

In the right panel of Figure \ref{Figure:Solutions} we plot instead $\Omega_{\rm eff}=(\overline{\dom{\mH}^2}-\mH^2)/\mH^2$. Qualitatively the behaviour with $\Rs$ is the same as that for the Hubble rate. Both $(\overline{\dom{\mH}}-\mH)/\mH$ and $\Omega_{\rm eff}$ are negative in this gauge and frame. To obtain $\Omega_{\rm eff}\lesssim 10^{-2}$ would require $\Rs\gtrsim 24{\rm Mpc}$.

The arbitrariness of the results is extremely unsatisfying and stems from the strong ultra-violet divergences. The automatic conclusion is that contrary to expectation this gauge in the gravitational frame is not well-suited to calculations of backreaction. Moreover, since $\Omega_{\rm eff}\equiv 0$ in the uniform curvature gauge, traceless uniform CDM gauge certainly can not be used to approximate the more well-defined choice of gauge. However, it is interesting to note that were we to take an infra-red divergence that appears in the domain volume seriously, and smooth the ultra-violet divergence in the integrals above, then the effective energy density would be driven to zero.

\subsection{Conformal Newtonian Gauge}
\noindent The final gauge we consider in the gravitational frame is conformal Newtonian gauge. With $C_{ij}=-\psi_N\delta_{ij}$, the average Hubble rate and effective energy density are
\bea
\overline{\dom{\mH}}&=&\mH+\frac{1}{3}\overline{\av{\dot{\psi}_{1N}+\frac{1}{2}\dot{\psi}_{2N}-6\psi_{1N}\dot{\psi}_{1N}}}=\frac{1}{6}\overline{\av{\dot{\psi}_{2N}}}, \\
\mH^2\Omega_{\rm eff}&=&\frac{2}{3}\mH\overline{\av{\dot{\psi}_{1N}+\frac{1}{2}\dot{\psi}_{2N}-6\psi_{1N}\dot{\psi}_{1N}}}+\frac{1}{9}\overline{\av{\dot{\psi}_{1N}}^2}=\frac{1}{3}\mH\overline{\av{\dot{\psi}_{2N}}},
\eea
where we have used that $\dot{\psi}_{1N}=0$ to simplify the forms considerably. Using the analytic expression for $\psi_{2N}$ in equation (\ref{AnalyticPsi2}) reduces this to
\be
\overline{\av{\dot{\psi}_{2N}}}=\frac{7}{30}\dot{B}_3(\eta)\ktwoint\approx \frac{1}{9}\eta\ktwoint
\ee
where we have used $\dot{B}_3\approx (10/21)\eta$. We therefore have the average Hubble rate and effective energy density
\be
\overline{\dom{\mH}}=\mH+\frac{1}{54}\eta\ktwoint, \quad
\mH^2\Omega_{\rm eff}=\frac{2}{27}\ktwoint .
\ee
These integrands are plotted in Figure \ref{Figure:Integrands} and contain neither infra-red nor ultra-violet divergences. Since the term of the form $\overline{\av{\cdot}\av{\cdot}}$ is vanishing in matter domination there is no dependence on the averaging domain; in a $\Lambda$CDM universe, however, this term would be non-vanishing and there would be a weak dependence on $\Rd$.\footnote{Strictly speaking this contradicts a statement in \cite{Umeh:2010pr} that the backreaction in gravitational frame does not depend on the choice of averaging domain. However, even in a $\Lambda$CDM universe $\dot{\psi}\ll\psi$ and the contribution will be entirely negligible, and for all practical purposes the conclusions of that paper are unchanged.}

It is also clear that in matter domination the effective energy density of the backreaction becomes a constant, in qualitative agreement with \cite{Brown:2009a,Clarkson:2009,Umeh:2010pr}.

From the form of the integrands the modifications to the Hubble rate in Newtonian gauge will be both significantly smaller than, and of an opposite sign to, those in uniform traceless CDM gauge. For the test EdS cosmology, we find
\be
\frac{\overline{\dom{\mH}}-\mH}{\mH}=3.44\times 10^{-7}, \quad
\Omega_{\rm eff}\approx 6.87\times 10^{-7} .
\ee
Using the zero-baryon transfer functions yields the slightly larger $(\overline{\dom{\mH}}-\mH)/\mH=4.4\times 10^{-7}$ and $\Omega_{\rm eff}\approx 9\times 10^{-7}$.

\section{The Projected Fluid Frame}
\noindent Averaging in the projected fluid frame tangles together fluid and metric quantities and it is not possible to significantly simplify expressions employing the dynamical constraints. In particular, the term $\mH_{\mathcal{D},1}=\av{\partial^a\partial_av+\dot{C}+2\mH C}$ is present at both linear and second-order in perturbations and does not readily simplify. For instance, at linear order the Laplacian of the velocity can be replaced with a combination of the lapse, shift and curvature using the momentum constraint $\dot{\psi}+\mH\phi=-4\pi Ga^2(v+B)$; removing the velocity then results in the average of a combination of $\phi$, $B$, $\psi$ and $E$, which will not cancel. The situation at second-order is significantly more complicated.

Uniform curvature gauge remains the uniquely well-defined gauge in which to perform a spatial average. It can then be argued that a calculation of the backreaction in this frame and gauge is the best we can hope for within the confines of cosmological perturbation theory.

In the gravitational frame, we found that the traceless uniform CDM gauge is not a good approximator for the uniform curvature gauge, despite being almost volume-preserving. In the fluid frame there is an extra motivation for studying this gauge: we can calculate the backreaction true to second order using only linear perturbations since $v_T=C_T=0$. If the uniform traceless CDM gauge is a reasonable approximation to the uniform curvature gauge, it is then significantly more straightforward to find results in this gauge.

\subsection{Uniform Curvature Gauge}
\noindent In uniform curvature gauge, the averaged Hubble rate (\ref{FluidFrameHubbleRate}) simplifies slightly to become
\be
\dom{\mH}=\mH+\frac{1}{3}\av{\partial^a\partial_a v_{1F}}+\frac{1}{6}\av{\partial^a\partial_a v_{1F}}+\frac{1}{3}\av{\phi_{1F}\partial^a\partial_av_{1F}-\partial_a\phi_{1F}\partial^aB_{1F}+\frac{3}{2}\mH\partial^aV_{1F}\partial_aV_{1F}} .
\ee
The ensemble average of the Hubble rate and the effective energy density are therefore
\bea
\overline{\dom{\mH}}&=&\mH+\frac{1}{3}\overline{\av{\partial^a\partial_av_{1F}+\frac{1}{2}\partial^a\partial_av_{2F}}}+\frac{1}{3}\overline{\av{\phi_{1F}\partial^a\partial_av_{1F}-\partial_a\phi_{1F}\partial^aB_{1F}+\frac{3}{2}\mH\partial^aV_{1F}\partial_aV_{1F}}}, \\
\mH^2\Omega_{\rm eff}&=&\frac{2}{3}\mH\overline{\av{\partial^a\partial_av_{1F}+\frac{1}{2}\partial^a\partial_av_{2F}}}
\nonumber \\ &&\quad
+\frac{2}{3}\mH\overline{\av{\phi_{1F}\partial^a\partial_av_{1F}-\partial_a\phi_{1F}\partial^aB_{1F}+\frac{3}{2}\mH\partial^aV_{1F}\partial_aV_{1F}}}+\frac{1}{9}\overline{\av{\partial^a\partial_av_{1F}}^2} .
\eea
We consider these averages term by term. Since the ensemble average of a linear perturbation vanishes, $\overline{\av{\partial^a\partial_av_{1F}}}=0$. From equation (\ref{vf2}) we can also see that
\bea
\overline{\av{\partial^a\partial_av_{2F}}}&=&\overline{\av{-\frac{10}{63}\eta^3\partial^i\partial_j\left(\partial_i\phi_0\partial^j\phi_0\right)+\frac{1}{21}\eta^3\partial^i\partial_i\left(\partial^a\phi_0\partial_a\phi_0\right)+\frac{1}{9}\eta^3\partial^i\left(\partial_i\phi_0\partial^a\partial_a\phi_0\right)}}
\\ && \quad
+\overline{\av{\frac{5}{4}\eta\chi_3(\phi_0)-\eta\phi_0\partial^a\partial_a\phi_0-\frac{17}{12}\eta\partial^a\phi_0\partial_a\phi_0}}
\eea
The first of these terms is
\be
-\overline{\av{\frac{10}{63}\eta^3\partial^i\partial_j\left(\partial_i\phi_0\partial^j\phi_0\right)}}=-\frac{10}{63}\eta^3\overline{\av{2\partial^i\partial_i\partial_j\phi_0\partial^j\phi_0+\partial_i\partial_j\phi_0\partial^i\partial^j\phi_0+\partial^i\partial_i\phi_0\partial^j\partial_j\phi_0}} .
\ee
Transferring this to Fourier space reveals that this term vanishes on ensemble averaging. The second term is
\be
\overline{\av{\frac{1}{21}\eta^3\partial^i\partial_i\left(\partial^a\phi_0\partial_a\phi_0\right)}}=\frac{1}{21}\eta^3\overline{\av{\partial^i\partial_i\partial_j\phi_0\partial^j\phi_0+2\partial_i\partial_j\phi_0\partial^i\partial^j\phi_0+\partial^i\partial_i\phi_0\partial^j\partial_j\phi_0}}
\ee
which is readily seen to also vanish on ensemble averaging. The third term,
\be
\overline{\av{\frac{1}{9}\eta^3\partial^i\left(\partial_i\phi_0\partial^a\partial_a\phi_0\right)}}=\frac{1}{9}\eta^3\overline{\av{\partial^i\partial_i\phi_0\partial^a\partial_a\phi_0+\partial_i\phi_0\partial^i\partial^a\partial_a\phi_0}}
\ee
also vanishes. In \cite{Clarkson:2009} it is shown that
\be
\overline{\av{\chi_3(\phi_0)}}=\frac{1}{3}\overline{\av{\partial^a\phi_0\partial_a\phi_0}}
\ee
and since
\be
\overline{\av{\partial^a\phi_0\partial_a\phi_0}}=-\overline{\av{\phi_0\partial^a\partial_a\phi_0}}=\ktwoint
\ee
we can see that
\be
\overline{\av{\partial^a\partial_av_{2F}}}=0 .
\ee
The ensemble averaged Hubble rate then reduces to
\be
\overline{\dom{\mH}}=\mH+\frac{25}{18}\eta\ktwoint .
\ee

In \cite{Clarkson:2009} it was additionally shown that
\be
\overline{\av{\partial^a\partial_aA}^2}=\int k^4\left|A(k)\right|^2W^2(k\dom{R})\frac{{\rm d}k}{k}
\ee
where $\dom{R}$ is a length scale characterising the averaging domain. Using this we can see that the effective energy density in uniform curvature gauge and the projected fluid frame is
\be
\mH^2\Omega_{\rm eff}=\frac{50}{9}\ktwoint+\frac{1}{18}\eta^2\kfourintw .
\ee
The only scale dependence in the solution enters in this final term -- for a large enough volume, the integral is driven to zero and the effective energy density is governed by the first term.

The right panel of Figure \ref{Figure:Integrands} shows the integrands of $\overline{\dom{\mH}}$ and of $\Omega_{\rm eff}$. We can see that the impacts on the Hubble rate and its square are both positive, 
are cleanly under control, and will be smaller than those in traceless uniform CDM gauge. Setting the domain scale to the Hubble scale gives
\be
\frac{\overline{\dom{\mH}}-\mH}{\mH}=1.8\times 10^{-4} ,
 \quad 
\Omega_{\rm eff}=3.61\times 10^{-4} .
\ee
Using the zero-baryon transfer function yields the slightly larger $(\overline{\dom{\mH}}-\mH)/\mH=2.3\times 10^{-4}$, $\Omega_{\rm eff}=4.6\times 10^{-4}$. This result is directly comparable with that in \cite{Umeh:2010pr} which performed the equivalent calculation in conformal Newtonian gauge, with the result $\Omega_{\rm eff}\approx 4\times 10^{-4}$, $(\overline{\dom{\mH}}-\mH)/\mH\approx 2\times 10^{-4}$ for an Einstein-de Sitter cosmology. It is also approximately in line with earlier calculations such as those in \cite{Behrend:2008,Paranjape:2008,Brown:2009a}.

\begin{figure}
\includegraphics[width=0.45\textwidth]{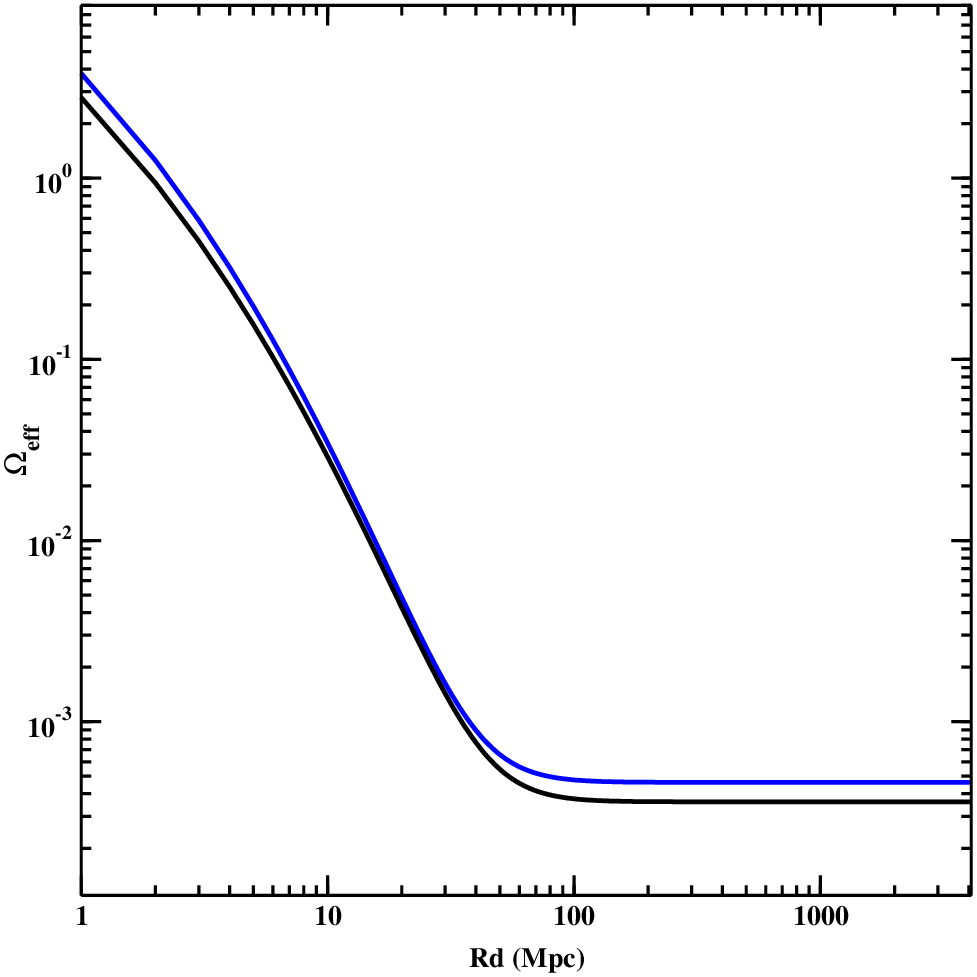}
\caption{Effective energy density of the backreaction in the projected fluid frame and uniform curvature gauge, as a function of domain radius $\dom{R}$ from accurate numerical calculations (black) and the zero-baryon approximation (blue).}
\label{RdH2}
\end{figure}

Figure \ref{RdH2} shows the dependence of the effective energy density on the averaging domain scale. At $\dom{R}=16{\rm Mpc}$ the effective energy density is $\Omega_{\rm eff}=8\times 10^{-3}$, decaying to $\Omega_{\rm eff}=4.4\times 10^{-4}$ at $\dom{R}=64{\rm Mpc}$. In this gauge it is then possible to identify a loose ``homogeneity scale'' at $\dom{R}\approx 150-250{\rm Mpc}$, above which the backreaction becomes scale-independent. This agrees well with the calculation in \cite{Clarkson:2009,Umeh:2010pr}, which identified a similar scale in conformal Newtonian gauge.

\subsection{Traceless Uniform CDM Gauge}
\noindent The traceless uniform CDM gauge comoves with the CDM, and is volume-preserving to first-order in perturbations. Selecting this gauge, the averaged Hubble rate is
\be
\dom{\mH}=\mH+\frac{1}{3}\av{\frac{3}{2}\mH\partial^aB\partial_aB-\partial^a\phi\partial_aB}-\frac{2}{3}\av{\dot{C}^{ab}C_{ab}}
\ee
and we can therefore calculate the result up to second-order in perturbation theory employing only linear perturbations. The second term is identical to that in the gravitational frame, and so
\be
\dom{\mH}=\mH+\mH_{\rm Grav.}+\frac{1}{3}\av{\frac{3}{2}\mH\partial^aB\partial_aB-\partial^a\phi\partial_aB} .
\ee
Using the perturbations given in (\ref{TracelessCDM}) we find that
\be
\overline{\av{\frac{3}{2}\mH\partial^aB\partial_aB-\partial^a\phi\partial_aB}}=\frac{\eta^5}{162}\int k^6\left|\phi_0\right|^2\frac{{\rm d}k}{k} .
\ee
This contains a severe ultra-violet divergence which we control with a smoothing scale $\Rs$. The severity of the divergence suggests we must smooth on much larger scales to control otherwise divergent results.

The averaged Hubble rate and effective energy density are then
\bea
\overline{\dom{\mH}}-\mH&=&\frac{1}{486}\eta^5\int k^6\mathcal{P}(k)\left|\phi_0\right|^2W^2(kR_S)\frac{{\rm d}k}{k}-\mH_{\rm Grav.}, \\
\mH^2\Omega_{\rm eff}&=&\frac{2}{243}\eta^4\int k^6\mathcal{P}(k)\left|\phi_0\right|^2W^2(kR_S)\frac{{\rm d}k}{k}-\mH^2\Omega_{\rm eff, Grav.} .
\eea
The integrands are plotted in Figure \ref{Figure:Integrands}. On superhorizon scales these agree with the traceless uniform CDM results in the gravitational frame, but on subhorizon scales the ultra-violet divergence is extremely notable. Even smoothed at $\Rs=6{\rm Mpc}$, it is clear that the results in the projected fluid frame will be orders of magnitude larger than those in the gravitational frame. There is also a sign change, suggesting that the sign of the backreaction, at least, will be in agreement with the more controlled calculations in uniform curvature and conformal Newtonian gauges. Evaluating the present-day backreaction at $\Rs=6{\rm Mpc}$,
\be
\frac{\overline{\dom{\mH}}-\mH}{\mH}=3.9\times 10^4, \quad
\Omega_{\rm eff}=7.82\times 10^4 !
\ee

The solutions as a function of $\Rs$, for $\dom{R}=\eta_0/2$, are presented in Figure \ref{Figure:Solutions}. As might be expected, for a sufficiently large $\Rs$ -- a smoothing scale approaching that of the Hubble scale itself, $\Rs\gtrsim 2000{\rm Mpc}$ -- the results in the two frames coincide with one-another. On smaller scales the effective energy density is indeed positive, but we require a smoothing scale $\Rs\gtrsim 60{\rm Mpc}$ if we want $\Omega_{\rm eff}\lesssim 1$ -- to ensure $\Omega_{\rm eff}\lesssim 0.01$ we need $\Rs\gtrsim 130{\rm Mpc}$! The extreme divergence, and the smoothing scales required to control it to recover meaningful results, suggest that the 3-surface and its perturbations in this gauge are badly-suited to the study of backreaction. It is certainly possible to argue that a comoving uniform density gauge is not well-adapted to the recent universe; the recent universe contains both large velocities and high densities, implying that the hypersurface and its embedding grow increasingly contorted. In any event it is certain that we cannot use the traceless uniform CDM gauge as a convenient substitute for the uniform curvature gauge. While in an EdS universe a choice of $\Rs\approx 243{\rm Mpc}$ recovers equivalent results, we can not expect the same to hold true in more realistic universes -- the smoothing scale required for consistency will not remain the same.

\section{Discussion}
\noindent In this paper we have shown that the cosmological backreaction is both highly gauge-dependent and highly frame-dependent. We have advocated the use of a (comoving) volume-preserving coordinate system as the system in which a spacetime average is well-defined, and argued that given the smallness of the tensor perturbations uniform curvature gauge provides a surface to average across which preserves the comoving volume, and in which the time-dependences in the averages cancel.

We then motivated an alternative choice of a comoving VPG, a gauge chosen to comove with surfaces of uniform cold dark matter density, with the spatial coordinates chosen such that the trace of the spatial metric vanishes. While this gauge does not preserve the comoving volume at second order in perturbation theory it does preserve it at linear order, which is adequate for calculations to second order in perturbation theory. A convenient feature of this gauge is that one can solve the backreaction to second order employing only linear perturbation theory.\footnote{Note, however, that this relies on it being possible to fix the gauge at second order in the same manner as we have at first order. While this seems likely, it has not been demonstrated.}

We compare both of these gauges against the conformal Newtonian gauge, employed in \cite{Kasai:2006,Tanaka:2007,Brown:2009a,Clarkson:2009,Umeh:2010pr}. In the gravitational frame, the backreaction in uniform curvature gauge vanishes identically. Since it is in the uniform curvature gauge that the backreaction should be defined, this suggests that the definition of the Hubble rate from the expansion of a 3-volume is too restrictive; with no reference to the fluid content of the universe it is also hard to recover meaning from the results. Further, while the backreaction in traceless uniform CDM gauge exhibits a strong dependence on choice of smoothing scale $\Rs$, results in the gravitational frame do not exhibit any dependence on the choice of averaging scale $\dom{R}$ which, as pointed out in \cite{Clarkson:2009,Umeh:2010pr}, is rather unnatural. The effective energy densities in the traceless uniform CDM and conformal Newtonian gauges possess different signs, and to ensure they are of equivalent size we must use a smoothing of the order of hundreds of megaparsecs. It is clear that it is not possible to use the traceless uniform CDM gauge as a simple alternative to uniform curvature.

In the projected fluid frame, the ultraviolet catastrophes in traceless uniform CDM gauge are exacerbated. To ensure $\Omega_{\rm eff}<1$, as seems reasonable, we must smooth perturbations on scales smaller than $\Rs=64{\rm Mpc}$! In contrast, in uniform curvature gauge we do not need to smooth perturbations on small scales, and we find $\Omega_{\rm eff}\approx 4\times 10^{-4}$. This is in line with the results of \cite{Clarkson:2009,Umeh:2010pr} in conformal Newtonian gauge, and also agrees with previous order-of-magnitude estimates such as in \cite{Brown:2009a}. The use of uniform curvature gauge does, however, require knowledge of fluid velocities at second-order in perturbation theory, which is in general non-trivial. 

The effective energy density in the projected fluid frame also exhibits the expected dependence on the averaging scale $\dom{R}$, with the impact decaying and asymptoting to a constant as $\dom{R}\rightarrow H_0^{-1}$. This is in agreement with the behaviour noted in \cite{Clarkson:2009,Umeh:2010pr}. In these papers it was also stated that the two frames should agree when the averaging scale is on the order of the Hubble scale. We confirm this for traceless uniform CDM gauge, but only if the ultra-violet smoothing scale is itself approaching the order of the Hubble scale.

In uniform curvature gauge, however, we find that the effective energy density tends towards a a constant, finite value in the projected fluid frame as the averaging radius grows to infinity, which contrasts with the identically vanishing result in the gravitational frame. In the gauge in which the averaging is properly defined, the results in the two gauges will never coincide no matter how large the averaging domain. This forces us to choose a frame in which to work; since it is defined from physical quantities that have meaning for an observer we advocate the use of the projected fluid frame.

We have presented a calculation of the cosmological backreaction in pure matter universes in the uniform curvature gauge, in which averaging is well-defined. The effective energy density of backreaction in this gauge agrees well with previous calculations in conformal Newtonian gauge. An alternative, which is well-defined to second-order in perturbations, does not provide consistent results. This gauge also ensures that corrections to the 3-volume remain formally small and that Taylor expansions remain valid, which is not the case in either of the alternative gauges. While the backreaction remains of order $10^{-4}-10^{-3}$, as in previous calculations, this value is now on a significantly firmer basis than before. The result is also slightly larger than the estimates in, for instance, \cite{Behrend:2008,Brown:2009a} and the calculations in \cite{Clarkson:2009,Umeh:2010pr,Kolb:2011zz}. While the direct impact from perturbation theory is still relatively minor, it is not so clear that it can simply be neglected. The present day universe is not well described by second-order perturbation theory. If perturbations induce backreactions at the order of $10^{-4}-10^{-3}$, larger inhomogeneous structures could be expected to have a larger impact -- conceivably of the order of $\gtrsim 10^{-2}$ and equivalent to the energy density of baryons themselves (see for instance \cite{Clarkson:2011uk,Kolb:2011zz,Buchert:2011sx} and \cite{Rasanen:2008}).

The result in a $\Lambda$ or $\phi$CDM universe will be somewhat less due to the washing out of structure from dark energy. Our result is then an upper limit on the present-day impact of second-order perturbations on the background. A more comprehensive study would require examination of the deceleration parameter or on other measures (such as the variance of the Hubble rate, considered for instance in \cite{Li08,Umeh:2010pr,Wiegand:2011je}). Further progress, valid in the present universe, will then likely require the study of fully non-linear solutions to GR.

\acknowledgments{IAB thanks David Wands, Tomi Koivisto, Karim Malik and Juliane Behrend for useful discussions. AAC acknowledges financial support from NSERC.}

\bibliography{Backreaction}

\begin{appendix}
\section{The Backreaction in the Buchert Approach}
\label{Buchert}
In this appendix we prove in the context of the standard ``Buchert'' approach that the 
gravitational frame backreaction vanishes identically in uniform curvature gauge. 
It is standard to connect the averaged Hubble rate to the fluid content of the universe. 
Applying the averaging procedure to the Hamiltonian constraint and evolution of 
the extrinsic curvature produces Friedmann-like (so-called ``Buchert'') equations 
in the domain (\cite{Brown:2009b}):
\bea
\label{BuchertHub}
\dom{\mH}^2&=&\frac{8\pi G}{3}\sum_f\av{\alpha^2\rho_{(f)}}+\frac{1}{3}\av{\alpha^2}\Lambda-\frac{1}{6}\left(\Rd+\Qd^T-6\sum_f\Fd^{(f)}\right), \\
\label{BuchertRay}
\frac{\dom{\ddot{a}}}{\dom{a}}&=&-\frac{4\pi G}{3}\sum_f\av{\alpha^2\left(\rho_{(f)}+3p_{(f)}\right)}+\frac{1}{3}\av{\alpha^2}\Lambda+\frac{1}{3}\left(\Pd^T+\Qd^T-3\sum_f\Fd^{(f)}\right) .
\eea
$\Qd^T$, $\Pd^T$, $\Rd$ and $\Fd$ are, respectively, the kinematic backreaction, dynamic backreaction, averaged curvature and fluid tilt, which corrects the fluid quantities between the surface orthogonal to $n^\mu$ and the rest-frame of a fluid with 4-velocity $u^\mu_{(f)}$, and are given in \cite{Brown:2009b}. $\rho_{(f)}$ is the rest-frame density of a fluid and $p_{(f)}$ its rest-frame pressure. The effective energy density is then
\be
\label{BuchertRhoEff}
\mH^2\Omega_{\rm eff}=\frac{1}{6}\left(6\sum_f\Fd^{(f)}-\Rd-\Qd^T\right) .
\ee
Using $B_i=\partial_iB$, $v^i_{(a)}=\partial^iv_{(a)}$, the correction terms become
\be
\begin{array}{c}
\Rd=0, \quad \Qd=-4\mH\av{\partial^i\partial_iB}+\av{(\partial^i\partial_iB)(\partial^j\partial_jB)-(\partial^i\partial_jB)(\partial_i\partial^jB)}, \\
\Td^{(f)}=\dfrac{8\pi G}{3}a^2\bkr_{(f)}\av{\delta_{(f)}+2\phi}+\dfrac{8\pi G}{3}a^2\bkr_{(f)}\av{2\phi\delta_{(f)}+\partial_iB\partial^iB+(1+w_{(f)})\partial_iV_{(f)}\partial^iV_{(f)}} \\
\Ld=\dfrac{2}{3}a^2\Lambda\av{\phi}+\dfrac{1}{3}a^2\Lambda\av{\partial^iB\partial_iB} .
\end{array}
\ee
Expanding the perturbations into first- and second-order components, the effective energy density (\ref{BuchertRhoEff}) becomes
\bea
\frac{8\pi G}{3}a^2\bkr_{\rm eff}&=&\av{\frac{8\pi G}{3}a^2\sum_f\bkr_{(f)}\left(\delta_{(f,1)}+2\phi_{(1)}+\frac{1}{2}\delta_{(f,2)}+\phi_{(2)}\right)
+\frac{1}{3}a^2\Lambda\left(2\phi_{(1)}+\phi_{(2)}\right)+\frac{1}{3}\mH\partial^i\partial_i\left(2B_{(1)}+B_{(2)}\right)}
\nonumber \\ &&
+\av{\frac{8\pi G}{3}a^2\sum_a\bkr_{(a)}\left(2\phi_{(1)}\delta_{(a,1)}+\partial^iB_{(1)}\partial_iB_{(1)}+(1+w_{(a)})\partial_iV_{(a,1)}\partial^iV_{(a,1)}\right)+\frac{1}{3}a^2\Lambda\partial^iB_{(1)}\partial_iB_{(1)}}
\nonumber \\ &&
-\av{\frac{1}{6}\left(\left(\partial^i\partial_iB_{(1)}\right)\left(\partial^j\partial_jB_{(1)}\right)-\left(\partial^i\partial_jB_{(1)}\right)\left(\partial_i\partial^jB_{(1)}\right)\right) }.
\eea
The second-order perturbed Hamiltonian constraint in uniform curvature gauge \cite{Christopherson:2011ra} is
\bea
\lefteqn{
2\mH\partial^i\partial_iB_{(2)}+6\mH^2\phi_{(2)}+8\pi Ga^2\sum_f\bkr_{(f)}\delta_{(f,2)}=}
\nonumber\\&&
-16\pi Ga^2\sum_f\bkr_{(f)}(1+w_{(f)})\partial^iV_{(f,1)}\partial_iv_{(f,1)}+4\mH\partial^iB_{(1)}\partial_i\phi_{(1)}+\left(\partial^i\partial_iB_{(1)}\right)\left(\partial^j\partial_jB_{(1)}\right)-\left(\partial^i\partial_jB_{(1)}\right)\left(\partial_i\partial^jB_{(1)}\right)
\nonumber\\&&
+6\mH^2\left(4\phi_{(1)}^2-\partial^iB_{(1)}\partial_iB_{(1)}\right)+8\mH\phi_{(1)}\partial^i\partial_iB_{(1)} .
\eea
Employing this to eliminate $\partial^i\partial_iB_{(2)}$ in the effective energy density, and using the Friedmann equation to absorb terms proportional to $\Lambda$ leads ultimately to
\bea
\label{RealSpaceEffectiveEnergy}
\lefteqn{
\mH^2\Omega_{\rm eff}=
\left<
\frac{8\pi G}{3}a^2\sum_f\bkr_{(f)}\delta_{(f,1)}+2\mathcal{H}^2\phi_{(1)}+\frac{2}{3}\mH\partial^i\partial_iB_{(1)}
\right>
}
\\ &&
+\left<\frac{8\pi G}{3}a^2\sum_f\bkr_{(f)}\left(2\phi_{(1)}\delta_{(f,1)}+(1+w_{(f)})\partial^iB_{(1)}\partial_iV_{(f,1)}\right)+4\mH^2\phi^2_{(1)}
+\frac{4}{3}\mH\phi_{(1)}\partial^i\partial_iB_{(1)}+\frac{2}{3}\mH\partial^i\phi_{(1)}\partial_iB_{(1)}\right>
\nonumber
\eea
The above form of the effective energy density contains only first-order perturbations, for which we have a complete and straightforward theory. In particular, we have the Hamiltonian and momentum constraints
\be
3\mH^2\phi_{(1)}+\mH\partial^i\partial_iB_{(1)}=-4\pi Ga^2\sum_f\bkr_{(f)}\delta_{(f,1)}, \quad
\mH\phi=-4\pi Ga^2\sum_f(1+w_{(f)})\bkr_{(f)}V_{(f,1)}
\ee
which can be used to easily eliminate the fluid quantities in terms of metric quantities. Use of the Hamiltonian constraint removes the average of first-order perturbations that appears in $\mH^2\Omega_{\rm eff}$, which becomes
\bea
\mH^2\Omega_{\rm eff}&=&\av{\frac{8\pi G}{3}a^2\sum_f\bkr_{(f)}\left(2\phi_{(1)}\delta_{(f,1)}+(1+w_{(f)})\partial^iB_{(1)}\partial_iV_{(f,1)}\right)
\right. \nonumber \\ && \left.
+4\mH^2\phi^2_{(1)}+\frac{4}{3}\mH\phi_{(1)}\partial^i\partial_iB_{(1)}+\frac{2}{3}\mH\partial^i\phi_{(1)}\partial_iB_{(1)}} .
\eea
The Hamiltonian and momentum constraints give
\bea
2\phi_{(1)}\frac{8\pi G}{3}a^2\sum_f\bkr_{(f)}\delta_{(f,1)}+4\mathcal{H}^2\phi^2_{(1)}+\frac{4}{3}\mH\phi_{(1)}\partial^i\partial_iB_{(1)}&=&0,
\nonumber \\
\partial^iB_{(1)}\frac{8\pi G}{3}a^2\sum_f(1+w_f)\bkr_{(f)}\partial_iV_{(f,1)}+\frac{2}{3}\mH\partial^iB_{(1)}\partial_i\phi_{(1)}&=&0 .
\eea
A quick examination of the effective energy density of the backreaction quickly confirms that
\be
\mH^2\Omega_{\rm eff}=0.
\ee
We have verified, within the full Buchert approach, that the gravitational-frame backreaction up to second-order in perturbation theory in uniform curvature gauge vanishes identically! Note that in forming this conclusion we have not transferred the system into Fourier space nor have we taken an ensemble average -- the conclusion follows inevitably, in real space, for scales on which second-order perturbation theory is valid, and for any admixture of fluids.

\end{appendix}

\end{document}